\documentclass[manuscript=article]{achemso}
\setkeys{acs}{doi = true}
\setkeys{acs}{journal = apchd5}
\usepackage{graphicx} 
\usepackage[T1]{fontenc}
\usepackage{siunitx}
\usepackage{xcolor}
\usepackage[normalem]{ulem}
\usepackage{amsmath}
\usepackage{amssymb}
\usepackage{setspace}
\usepackage{wrapfig}
\input{insbox}
\makeatletter
\renewcommand*{\acs@author@fnsymbol@symbol}[1]{%
\ifcase #1 *\or
    1\or
    2\or
    3
    \fi
}

\setcitestyle{numbers,round}
 \setkeys{acs}{keywords = true}

\DeclareSIUnit\angstrom{\mbox{\normalfont\AA}}
\DeclareSIUnit\bar{\mbox{\normalfont bar}}

\newcommand\fig[1]{Fig.~\ref{fig:#1}}
\newcommand\figs[1]{Figs.~\ref{fig:#1}}

\newcommand\wse{WSe$_2$}

\usepackage[hidelinks,hyperfootnotes=false]{hyperref}
\title{Photoemission electron microscopy of exciton-polaritons in thin \wse{} waveguides}

\author{Tobias Eul}
\email{eul@physik.uni-kiel.de}
\author{Miwan Sabir}
\author{Victor DeManuel-Gonzalez}
\author{Florian Diekmann}
\affiliation{Institute of Experimental and Applied Physics, Kiel University, 24098 Kiel, Germany}
\author{Kai Rossnagel}
\affiliation{Institute of Experimental and Applied Physics, Kiel University, 24098 Kiel, Germany}
\alsoaffiliation{Kiel Nano, Surface and Interface Science KiNSIS, Kiel University, 24118 Kiel, Germany}
\author{Michael Bauer}
\affiliation{Institute of Experimental and Applied Physics, Kiel University, 24098 Kiel, Germany}
\alsoaffiliation{Kiel Nano, Surface and Interface Science KiNSIS, Kiel University, 24118 Kiel, Germany}

\date{\today}

\begin{document}
\singlespacing


\begin{abstract}
\InsertBoxR{0}{\includegraphics[width=0.4\textwidth]{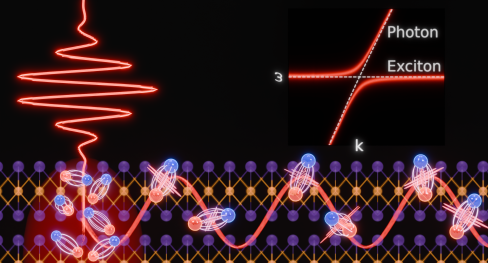}}
Exciton-polaritons emerging from the interaction of photons and excitons in the strong coupling regime are intriguing quasiparticles for the potential exchange of energy during light-matter interaction processes such as light harvesting.
The coupling causes an energy anti-crossing in the photon dispersion centered around the exciton resonance, i.e., a Rabi splitting between a lower and upper energetic branch.
The size of this splitting correlates with the coupling strength between the exciton and the photonic modes.
In this work, we investigate this coupling between excitons and photonic waveguide modes excited simultaneously in thin-film flakes of the transition-metal dichalcogenide \wse.
Using a Photoemission electron microscope, we are able to extract the dispersion of the transverse electric and magnetic modes propagating through these flakes as well as extract the energy splitting.
Ultimately, our findings provide a basis for the investigation of the propagation of exciton-polaritons in the time-domain via time-resolved photoemission.
\end{abstract}

\maketitle



\section{Introduction}
Excitons, bound electron-hole pairs, are the focus of numerous experimental and theoretical research efforts based on their potential for light-driven energy harvesting. 
From the many semiconducting materials exhibiting excitonic resonances, transition-metal dichalcogenides (TMDC) have garnered special interest in recent years since they exhibit excitons with large binding energies \cite{Cheiwchanchamnangij.2012,Chernikov.2014,Zhu.2015,Wang.2018}.
Embedding TMDC films or even monolayers within micrometer-sized cavities creates a suitable environment for a strong interaction between excitons and photons \cite{Dufferwiel.2015,Liu.2015,Lundt.2016, Flatten.2016, Liu.2017, Chen.2017}.
This interaction facilitates the formation of the light-matter quasiparticle termed exciton-polaritons \cite{Weisbuch.1992,Gibbs.2011}. 
Its hybrid nature promises new avenues for the energy exchange between light and matter and thus new technological developments.

\begin{figure}[!b]
  \includegraphics{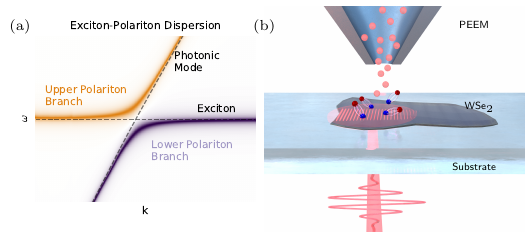}
  \caption{\textbf{Exciton-polariton dispersion and experimental scheme.}
  (a) Schematic illustration of the exciton-polariton dispersion.
  (b) Experimental setup with thin \wse{} flakes situated on a transparent glass substrate.
  A laser illuminates the flake from the bottom, where it excites the exciton-polaritons as well as generates the photoelectrons, which enter the PEEM lens system.
  }
  \label{fig:intro}
\end{figure}

The transfer of energy via the propagation of such a quasiparticle is especially interesting.
Therefore, the interaction of excitons with propagating light modes rather than spatially confined cavity photons may provide another suitable environment for exciton-polaritons.
Essentially, thin films of TMDC materials themselves host the necessary waveguiding modes, which are able to interact with the intrinsic exciton resonances  \cite{Zong.2023}.
Recent near-field studies involving scanning-near field microscopy (SNOM) \cite{Fei.2016,Hu.2017} as well as cathode luminescence (CL) \cite{Taleb.2022} focused on this particular approach for exciton-polaritons and discussed their spectroscopic signatures in thin films of \wse{} and MoSe$_2$.
The dispersion of these polaritons exhibits a distinct energy anti-crossing centered around the exciton resonance, as illustrated in \fig{intro}(a).

Photoemission electron microscopy (PEEM) is another experimental technique that demonstrates significant sensitivity to the electric near-field and surface waves in both metallic \cite{Kubo.2005,Heringdorf.2007,Leissner.2012,Kahl.2014,Spektor.2017,Joly.2018,Dabrowski.2020,Großmann.2021,Timothy.2020, Kilbane.2023} and photonic materials \cite{Fitzgerald.2013, Word.2013, Fitzgerald.2014, Aeschlimann.2015, Klick.2018, Stenmark.2019, Aeschlimann.2023}.
Furthermore, it is possible to measure dispersion relations of these propagating modes by tuning the wavelength of the exciting laser \cite{Fitzgerald.2013,Leissner.2013,Lemke.2014,Hartelt.2021}. 
The direct imaging of the surface in PEEM in combination with pump-probe excitation schemes additionally facilitates tracing the propagation of waves and quasiparticles along the surface in the time domain \cite{Kubo.2007,Lemke.2012, Spektor.2017, Kahl.2018}.
This makes it a promising method for tracing the propagation of exciton-polaritons in real-time. 
Ultimately, time-resolved PEEM may even reveal highly local information on the energy transfer, i.e., the Rabi oscillation, between exciton and waveguide mode on an ultrafast timescale.

\section{Results and Discussion}
In this work, we demonstrate the capability of the PEEM technique to measure the dispersion relation of exciton-polaritons in thin \wse{} films and determine the coupling strength between the A-exciton and photonic waveguide modes from this data.
\fig{intro}(b) depicts our experimental geometry.
A laser with a tunable center wavelength in the range of \SIrange{695}{900}{\nano\meter} corresponding to photon energies $E_{\mathrm{Ph}}$ between \SI{1.38}{\electronvolt} and \SI{1.78}{\electronvolt} illuminates the TMDC sample from the bottom, and allows us to spectrally probe the dispersion around the A exciton in \wse{} at $E_\mathrm{ex}=\SI{1.6}{\electronvolt}$ \cite{Fei.2016,Taleb.2022}.
A high electric field accelerates the photoelectrons carrying the information on the quasiparticle interactions into the PEEM column. 
The work function of \wse{}, which must be overcome in the photoemission process, is in the order of \SI{4.3}{\electronvolt} \cite{Kim.2021} and is considerably higher than our photon energies.
We reduce the work function of our samples using a procedure that is customary for the PEEM-method \cite{Cinchetti.2003, Aeschlimann.2007,Lemke.2014,Timothy.2020,Hartelt.2021}: We evaporate a fraction of a monolayer of an alkali metal (K) on the surface until it is low enough to enable the emission of photoelectrons via two-photon photoemission (2PPE).
In this limit, we are able to generate photoelectrons with the same laser pulses that excite the exciton-polariton.
In principle, our study would also be possible without lowering the work function, using three-photon photoemission.
We discuss the substantial limitations of this approach in the Supplementary information (SI).

\begin{figure*}[t]
   \includegraphics[width=\textwidth]{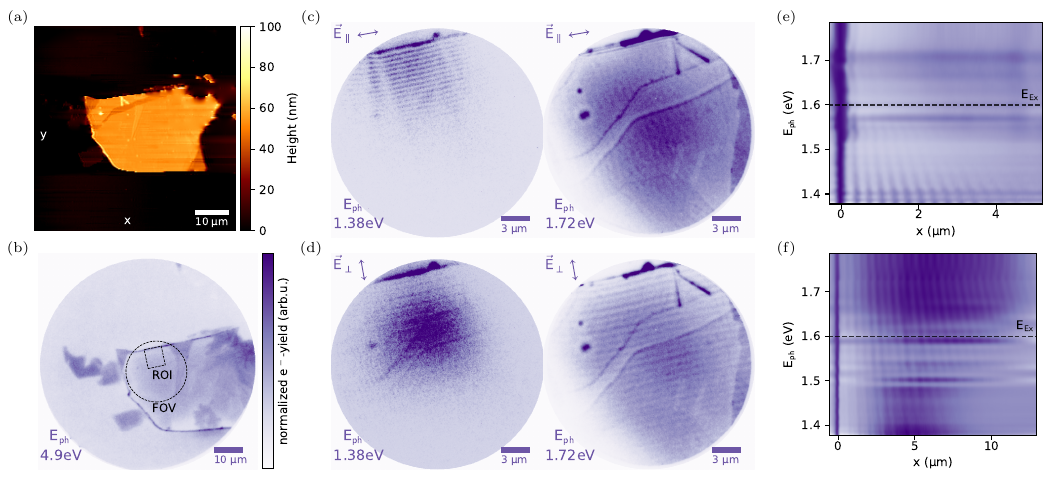}
   \caption{\textbf{Waveguide mode excitation in \wse{} investigated using PEEM.}
   (a) AFM image of a \wse{} flake with a thickness of approximately \SI{55}{\nano\meter}.
   (b) PEEM image of the same flake recorded with a photon energy of \SI{4.9}{\electronvolt}.
   The dashed circle (FOV) indicates the field of view of the detailed measurement in (c,d).
   The dashed square (ROI) indicates the region of interest for the extraction of the wave patterns. 
   (c,d) PEEM images recorded with photon energies of \SI{1.38}{\electronvolt} (left) and \SI{1.72}{\electronvolt} (right) with the polarization parallel and perpendicular to the upper edge of the flake, respectively.
   (e,f) Extracted wave patterns for parallel and perpendicular polarization as a function of excitation photon energy, respectively.
   The dashed line indicates the energy $E_{\mathrm{Ex}}$ of the A-exciton of \wse{}.
   }
   \label{fig:Waves}
\end{figure*}

We prepared the \wse{} flakes on indium-tin-oxide (ITO) coated glass substrates via the mechanical exfoliation method.
This method yields flakes with clean, flat surfaces and well defined edges albeit with varying thicknesses (\SIrange{20}{150}{\nano\meter}) and lateral dimensions (\SIrange{5}{50}{\micro\meter}).
Prior to the PEEM experiments, we measured the thickness of the flakes with an atomic force microscope (AFM).
\fig{Waves}(a) shows the AFM image for an exemplary \wse{} flake, which exhibits a thickness of approximately \SI{55}{\nano\meter} as well as sizes of \SI{30}{\micro\meter} and \SI{25}{\micro\meter} in x- and y-direction, respectively.
A more detailed description of our methods follows the main text below.

The corresponding PEEM image in \fig{Waves}(b), recorded with a mercury discharge lamp at a photon energy of $E_\mathrm{ph}=\SI{4.9}{\electronvolt}$, shows the topography based on work function differences of the substrate and the \wse{} flake.
The dashed circle (FOV) in \fig{Waves}(b) highlights the field of view, where we zoomed in on the sample to get close up images of the optical near-field of the waveguide modes.
The dashed square (ROI) on the other hand marks the region of interest used for the subsequent quantitative analysis.
For the experiment, we focus our laser onto the edge of the flake visible at the border of the ROI.
Here, the light can couple into different waveguide modes, depending on the in-plane polarization with respect to this edge.

A guided mode propagates through a planar waveguide by internal reflection upon the top and bottom interfaces.
The polarization dependent phase change upon this reflection determines the character of a mode in the waveguide.
Considering the two typical cases of excitation with s- and p-polarized light with respect to the waveguide geometry, two types of modes can be excited in a planar waveguide.
S-polarized light couples to transverse electric (TE) modes, where the electric field is perpendicular to the propagation direction.
P-polarized light couples to transverse magnetic (TM) modes, where the magnetic field is perpendicular to the propagation direction.
Since the wave propagates perpendicular to the excitation edge, we couple into a TE-mode for a laser polarization parallel to the edge.
Similarly, the laser polarization perpendicular to the edge excites a TM-mode.

When these waveguide modes propagate through the flake, its electric field interferes with that of the second photon in the 2PPE-process resulting in a wave pattern in the photoemission contrast on the surface \cite{Buckanie.2013,Kahl.2014}.
To investigate the dispersion relation of these modes,  we excite the flake edge with varying photon energies around the exciton resonance and examine the photoemitted wave patterns.

\figs{Waves}(c,d) show exemplary images for the photon energies $E_\mathrm{ph}=\SI{1.38}{\electronvolt}$ and $E_\mathrm{ph}=\SI{1.72}{\electronvolt}$, respectively, recorded with the polarization parallel (top) and perpendicular (bottom) to the edge.
In the case of parallel polarization, i.e. for excitation of a TE-mode, we see a distinct wave pattern for the low photon energy, whereas just one single period of a wave is visible for the large photon energy.
This result was to be expected based on the absorption properties of \wse{} above the energy $E_{\mathrm{Ex}}\approx\SI{1.6}{\electronvolt}$ of the A-exciton in \wse{} \cite{Fei.2016}.
This absorption influence should be even more apparent in our experiment due to the nonlinear 2PPE excitation scheme.
The wave patterns also clearly exhibit different periods hinting at the dispersion of the photonic mode.
For excitation of a TM-mode on the other hand, the wave pattern remains observable for both photon energies.
\figs{Waves}(e,f) illustrate this more clearly in a comparison of the wave patterns at all measured photon energies.
For the data comparison, we sum up the intensity in the ROI of the PEEM image along the direction of the edge.
As expected, the waves for the TE-mode in (e) show a decreasing propagation length with increasing photon energy.
Surprisingly, in the case of the TM-mode in (f) the propagation length does not differ significantly over the measured spectral region.
Since the exciton resonance is largely responsible for the absorption in this spectral region, it appears as if the waveguide mode excited via perpendicular polarization does not interact with the exciton, at least for this particular thickness of the flake.
This observation will be discussed further below.

\begin{figure*}[t]
    \includegraphics[width=\textwidth]{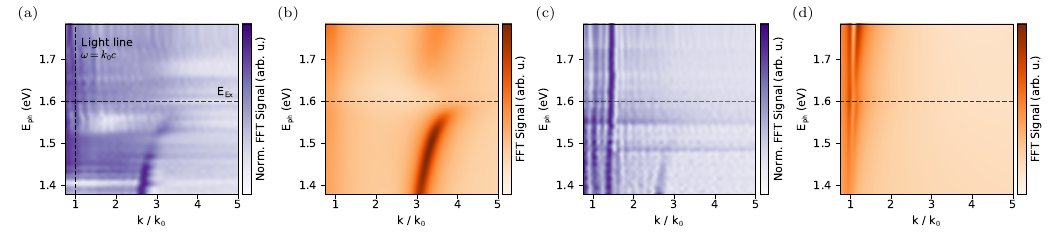}
    \caption{\textbf{Waveguide mode dispersion.}
    Dispersion curves extracted from the Fourier transformation of measured and simulated wave patterns.
    The wave numbers $k$ are relative to the wave number of light in vacuum $k_0$.
    (a,c) Experimental data for TE- and TM-mode, respectively.
    (b,d) FDTD simulations TE- and TM-mode, respectively.
    Simulated data for a flake thickness of \SI{55}{\nano\meter}.
    }
    \label{fig:Dispersion}
\end{figure*}
The next step of the analysis is to directly extract the dispersion of the photonic modes from the wave patterns.
By applying the Fast Fourier Transform (FFT) method to the wave patterns, we can analyze their periods in terms of the amplitude in Fourier space depending on the wavenumber $k$ relative to the wavenumber $k_0$ of light in free space.
Since the wave patterns are not ideal signals with decay and constant offset, we apply a high-pass filter prior to the FFT, which removes low-frequency contributions from the signal.
\fig{Dispersion} shows the results of this method in dispersion plots consisting of the extracted Fourier signal after normalization to its respective maximum for each photon energy and subsequent interpolation.
The normalization is necessary, since the experiment conditions such as focal spot size, intensity and, most importantly for the 2PPE process, the pulse durations vary for each individual photon energy.
Images (a,c) show the experimental data obtained from the wave patterns in \figs{Waves}(e,f), respectively.
The dispersion for the TE-mode in (a) clearly shows one of the characteristic signatures of strong coupling between exciton and photonic mode, i.e., the lower polariton branch bending in the vicinity of the exciton resonance.
The upper branch is not visible due to the strong absorption in this spectral region.

Examining the result for the TM-mode in (c), we first of all notice the same feature as in (a), albeit with a low intensity.
This stems from the laser polarization being not perfectly perpendicular to the edge of the \wse{} flake.
More importantly, we see a linearly dispersing feature that does not show any deviation from its straight line around the exciton resonance.
As already indicated above, this mode obviously does not interact with the exciton.

To explain our observations, we additionally simulated the experiment by means of the finite-difference time-domain method \cite{Tidy3d} and using the optical constants for \wse{} from Ref.~\cite{Munkhbat.2022}.
We use a broadband gaussian source, which covers the spectral range of the experimental data, and focus it on an edge of a thin film of \wse.
Equivalently, we extract the electric field wave patterns from a frequency monitor situated in the \wse{} film and subsequently apply the FFT method.
Here, we forego the normalization of the Fourier amplitude for each individual photon energy to retain the information about the absorption.
Images (b,d) show the simulated dispersions corresponding to the experimental data in (a,c), i.e., a thickness of \SI{55}{\nano\meter} and $E_\parallel$, $E_\perp$, respectively. 
They reproduce the experiment reasonably well, albeit with a difference in the absolute values of $k/k_0$,  which is probably due to differences in the optical constants for the simulations and the actual sample.
For an excitation with $E_\parallel$ in (b), the lower branch disperses similarly to the one obtained from the experiment.
However, here we can additionally see the upper branch with a significantly reduced intensity and larger linewidth than the lower branch.
The analysis of the individual electric field components (see Fig.~S3 in the SI) confirms our initial assignment between laser polarization and the two mode types within the waveguide.
The largest contribution to the total electric field, and therefore the dispersion signal, comes from the component parallel to the excitation edge.
Since the waveguide mode propagates in the direction perpendicular to the edge, we can attribute the dispersive feature to a transverse electric (TE) mode.
Similarly, we determine from the simulation data in \fig{Dispersion}(d) and Fig.~S3 that the light with $E_\perp$ indeed couples into a transverse magnetic (TM) mode.
As in the experiment, the dispersion in this case is a straight line, again without any sign of interaction with the exciton.
The experimental data and simulation in \fig{Dispersion}(c) and (d) also exhibits another feature at $k/k_0=1$, i.e., a mode that follows the dispersion of light in free space.
This additional feature stems from the excitation source.

Additional simulations in the SI and similar results in Ref.~\citenum{Fei.2016} show that for increasing flake thickness the exciton also starts to couple to the TM-mode.
Interestingly, for both the TE- and the TM-mode, there seems to be a threshold for the waveguide thickness that determines the coupling to the exciton.
For the TE-mode this thickness threshold lies between \SI{20}{\nano\meter} and \SI{30}{\nano\meter}, whereas the waveguide needs to be thicker than \SI{60}{\nano\meter} for the TM-mode to couple.
When analyzing the electric field localization within the waveguide (see Fig.~S5 and Fig.~S7), it becomes clear that the exciton coupling depends on how strongly the light is confined within the waveguide.
In the case of the TE-mode, the waveguides simply become too thin.
The electric field is still very localized, but too much of the light resides outside of the waveguide below \SI{30}{\nano\meter}.
For the TM-mode however, the electric field amplitudes starts to delocalize for thicknesses below \SI{60}{\nano\meter}, which is why the threshold is different from the TE-mode.

To summarize, the data show that for this particular thickness of the \wse{} flake, we are able to control the exciton-photon coupling with the laser polarization being the experimental control parameter.
We can either couple into the TE-mode for strong coupling or into the TM-mode for a freely propagating waveguide mode.

\begin{figure}[t]
    \includegraphics{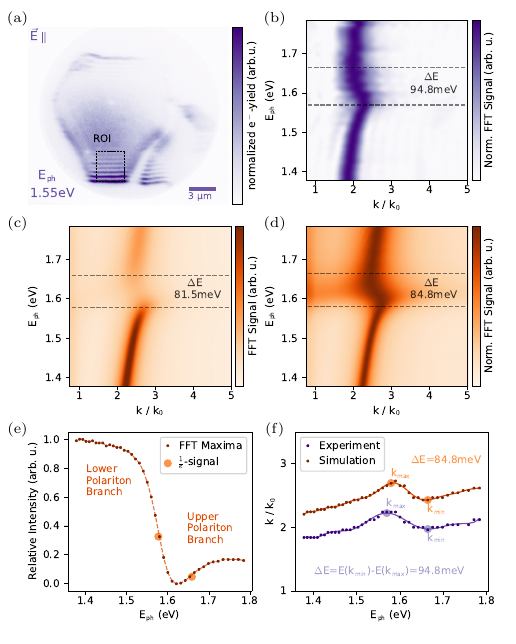}
    \caption{\textbf{Strong-coupling energy splitting.}
    (a) PEEM image of a \wse{} flake recorded with a photon energy of \SI{1.55}{\electronvolt} and laser polarization oriented parallel to the bottom edge.
    The dashed square (ROI) indicates the region of interest for the extraction of the wave patterns.
    (b) Dispersion of the TE-mode evaluated from the experimental data.
    (c,d) Simulated dispersion for a thickness of \SI{30}{\nano\meter} without and with normalizing the signal of each individual frequency.
    (e) Extracted maximum FFT signals from (c) for determining the energy splitting based on exponential reduction of the branch intensities.
    (f) Extracted maxima positions from (b) and (d) for determining the energy gaps based on turning points in the dispersion.
    }
    \label{fig:Splitting}
\end{figure}

Ultimately, we are interested in the coupling strength of the exciton-polariton, which can be determined from the energy splitting in the dispersion curve.
For this, we examine another \wse{} flake with a significantly smaller thickness so that the absorption is overall reduced.
As mentioned above, in this thickness range only the TE-mode couples to the exciton.
\fig{Splitting}(a) shows the PEEM image for this flake recorded at $E_\mathrm{ph}=\SI{1.55}{\electronvolt}$ with the polarization of the laser oriented parallel to the excitation edge at the bottom.
The dashed square (ROI) marks the region of interest for the analysis of the wave patterns.
Since the absorption is now reduced, we can resolve both the lower and the upper branch of the exciton-polariton in the dispersion curve in \fig{Splitting}(b).
Due to the individual normalization of the signal for each photon energy, we do not see a gap around the exciton resonance.
Nevertheless, we can assign an energy splitting between the two branches based on the turning points within the two polariton branches.
This becomes more evident when analyzing the corresponding FDTD simulations for a thickness of \SI{30}{\nano\meter} in \figs{Splitting}(c,d).

The two simulated dispersion plots show the same dispersion data without normalization in (c) and with normalization of the signal for each photon energy equivalent to the experimental data in (d).
From the data without normalization, it is possible to extract the energy splitting in a straightforward manner by analyzing the intensity distribution of the two branches.
\fig{Splitting}(e) shows the relative maximum intensity of the branches extracted for each photon energy.
In both branches we determine the point where the intensity has decreased to the $\frac{1}{e}$-fraction. 
We take the energy difference between these two points as a measure for the energy splitting of the dispersion curve.
For the examined flake thickness, the analysis yields an energy splitting of $\Delta E=\SI{81.5}{\milli\electronvolt}$.
The dashed lines in \fig{Splitting}(c) mark the extracted limits and the splitting.
For the normalized data, we evaluate the energy splitting based on the two turning points in the dispersion curve located between the two branches that are clearly visible in \fig{Splitting}(d).

For the quantitative analysis, we extract the $k/k_0$-values of the maximum in the dispersion for each photon energy and approximate this $k(E_\mathrm{ph}$) relation by applying a Savitzky-Golay filter on the data points.
\fig{Splitting}(f) shows this relation in orange for the simulated data.
The curve shows a clear local maximum and minimum, which we identify as the $k/k_0$-values of the turning points.
The corresponding photon energies at these points yield a value for the energy splitting of $\Delta E=\SI{84.4}{\milli\electronvolt}$.
This value is very close to the one extracted from the unnormalized data, which validates this approach.
We therefore apply the equivalent procedure to the experimentally determined dispersion curve.
The data from this analysis, shown as a  blue line in \fig{Splitting}(f), yields an energy splitting of $\Delta E=\SI{94.8}{\milli\electronvolt}$ for the exciton-polariton.

This value is in good agreement with the results of a SNOM study on exciton-photon coupling in MoSe$_2$ flakes \cite{Hu.2017}.
Conversely, the above mentioned CL study on \wse{} flakes \cite{Taleb.2022} discusses larger values of up to \SI{300}{\milli\electronvolt}.
This discrepancy stems from two effects present in the CL experiment.
Firstly, the excitation with the electron beam in the CL experiment results in much higher exciton densities in the interaction volume with the waveguide mode compared to the light induced excitation used in this study, which enhances the exciton-polariton coupling \cite{Wu.2023}.
Secondly, electron beams with sufficiently high beam energy generate Cherenkov radiation, which further increases the interaction strength with the excitons \cite{Chahshouri.2022}.

\section{Conclusion}
In summary, we studied and analyzed the coupling strength between photonic waveguide modes and the A-exciton in thin-film \wse{} flakes.
By varying the photon energy as well as the polarization of the exciting laser, we were able to extract the dispersion relation of different modes from the near-field wave patterns recorded with PEEM.
Additional FDTD simulations reproduced the experimental dispersion curves and provided us another avenue to analyze the strong coupling and the associated energy splitting between the lower and upper polariton branch.
The simulations affirmed the nature of the modes, i.e., TE or TM for a polarization parallel or perpendicular to the excitation edge, respectively.
With our experiment, we ascertained the challenge in accessing the upper polariton branch using nonlinear photoemission due to a strong absorption in \wse{} for this spectral region.
Nevertheless, we were able to extract the energy splitting from the turning points in the normalized dispersion curve, with $\Delta E$ in the order of \SI{100}{\milli\electronvolt} for the exciton-polariton formed from the TE-mode.

In comparison, PEEM provides similar information as the scanning near-field microscopy technique employed in previous work on exciton-polaritons \cite{Fei.2016,Hu.2017}.
For example, the extracted energy splitting compares well to the splitting of \SI{100}{\milli\electronvolt} observed with SNOM in thin films of MoSe$_2$.
While SNOM allows to perform experiment under ambient conditions, PEEM, in combination with varying the laser polarization, is more flexible in selectively exciting the different TE or TM modes.
Additionally, the parallel imaging of the surface, in contrast to scanning over it, allows for a faster data acquisition, and therefore facilitates an easier combination with pump-probe techniques for ultrafast time-domain studies.
We expect that time-resolved PEEM will provide a new avenue to investigate the propagation of exciton-polaritons as well as the periodic energy exchange between the involved quasiparticles in real time.




\section*{Experimental Details}
\subsection*{Photoemission electron microscope}
The photoemission electron microscope (Focus GmbH) consists of an electrostatic lens system, which images the lateral distribution of photoelectrons emitted from the sample surface.
An extractor lens situated directly in front of the sample collects the photoemitted electrons with a high voltage electric field, e.g. $V_\mathrm{ext}=\SI{10}{\kilo\volt}$ for this study.
The detector is a combination of a multichannel-plate (MCP), which multiplies the photoelectrons, and a phosphor screen, whose illumination is captured by a CCD-Camera (LaVision).
The photoelectrons are being detected in an energy integrated manner with a high spatial resolution of $\Delta x\approx\SI{50}{\nano\meter}$.
The overall system is contained within an ultrahigh vacuum chamber with pressures typically in the order of $p\leq\SI{5e-10}{\milli\bar}$.
The system additionally features a wire shaped alkali metal dispenser (SAES Getters) for the evaporation of small doses of potassium onto sample surfaces.

For the photoexcitation, the sample can be illuminated by laser sources through two different vacuum chamber entrance windows, either under a grazing incidence of \SI{65}{\degree} or under normal illumination from the back side of the sample.
The latter case additionally allows a very tight focusing of the laser beam down to \SI{5}{\micro\meter} spot sizes via a lens in close proximity to the sample.
More information on this normal incidence configuration can be found in \cite{Klick.2019}.
Additionally, the PEEM is equipped with a mercury discharge lamp (Focus GmbH) providing continuous wave, unpolarized light with a photon energy of \SI{4.9}{\electronvolt}, which is primarily used for overview images of the sample surface and adjusting based on work function contrasts.

\subsection*{Laser system}
Our laser system is a pulsed Titanium-Sapphire (Ti:Sa) Oscillator system (Tsunami, Spectra Physics) operated at a repetition rate of \SI{80}{\mega\hertz} and pumped with $P_\mathrm{pump}=\SI{7.5}{\watt}$ of the second harmonic of a diode pumped Neodymium-Yttrium Vanadate laser (Millennia Pro 10s, Spectra Physics).
The central wavelength of the Ti:Sa output is tunable in the range of \SIrange{700}{900}{\nano\meter}.
Depending on the output spectrum, the duration of the outgoing laser pulses varies in the range of \SIrange{100}{130}{\femto\second}.
Similarly, the output power lies within the range of $P_\mathrm{out}=$ \SIrange{0.5}{1}{\watt}.
Between the laser system and the PEEM no additional pulse compression was used because the priority of the experiment was a stable pointing of the beam onto the sample while tuning the central wavelength of the laser.
Two lenses used as a telescope, the vacuum chamber entrance window, an in-vacuum lens and a half wave plate are present in the beam path and lengthen our pulse duration upon beam transmission.
All reflective optics used were optimized for minimal group delay dispersion.

\subsection*{Alignment of sample and laser}
The half wave plate is situated directly before the vacuum chamber entrance window to control the in-plane polarization of the laser upon illuminating the sample under normal incidence.
The alignment of the laser with respect to the \wse{} flake is very crucial due to the rather small spot sizes of the laser in the order of \SIrange{5}{20}{\micro\meter} at the sample.
Ideally, the center of the beam should be on the excitation edge of the flake, to achieve the best coupling to the waveguide mode. 
Unfortunately, this would result in a very strong overexposure of the edge in the photoemission signal and very low contrast in the rest of the field of view.
Therefore, we position the spot in a way that enough of the gaussian profile clips the excitation edge to launch the waveguide mode, while the intensity maximum is in the center of the field of view.
This skews the overall intensity of the wave and makes it appear as if the waves decay away from the center of the field of view.
Depending on the photon energy this effect is more or less pronounced as evident from \fig{Waves}.
This is why we apply a high pass Fourier filter to the PEEM images along the propagation direction and normalize the data for each photon energy individually.

\subsection*{Sample fabrication}
Single crystals of 2Hb-WSe$_2$ were grown by iodine vapor transport.
High-purity tungsten wire and selenium pellets were sealed at one end of evacuated quartz ampoules (inner diameter \SI{12}{\milli\meter}, length \SI{200}{\milli\meter}) in a near stoichiometric ratio with approximately \SI{15}{\percent} selenium excess (\SI{3}{\milli\gram\per\centi\meter^3}) together with iodine granules (\SI{5}{\milli\gram\per\centi\meter^3}).
The ampoules were subjected to a temperature gradient of \SI{60}{\celsius}, with the hotter end containing the starting materials at \SI{900}{\celsius}.
The growth time for crystals forming near the colder end was \SI{900}{\hour}. 
The size of the crystals were up to 5x5~mm$^2$ in area and \SI{0.1}{\milli\meter} thick.
The doping of the semiconducting samples is typically p-type \cite{Buck.2011}.
We prepare our samples via mechanical exfoliation of the bulk WSe$_2$ crystal.
For this, we press a semiconductor wafer dicing tape on top of the bulk and remove the topmost layers.
With a second strip of this tape, we repeat the process on the material stuck on the first tape and afterwards the same with a third tape.
Finally, we press the third tape on top of our glass substrate to transfer the WSe$_2$ layers.
The glass substrate (Präzisions Glas \& Optik GmbH) is a standard soda-lime-glass with a thickness of \SI{1.1}{\milli\meter} covered with a few nano meters of a SiO$_2$ passivation layer.
On top of this is a sputter deposited thin film of indium-tin oxide with a thickness of  \SI{21}{\nano\meter} and a resistance of \SI{100}{\ohm}.
Prior to the PEEM experiments, we evaporate a fraction of a monolayer of potassium onto the sample surface to reduce the work function.
For this, we operate the K-dispenser at $I=\SI{5.7}{\ampere}$ for \SI{60}{\second} from a distance of approximately \SI{5.5}{\centi\meter} to the sample.
Over time and especially after illumination with the laser, the potassium level on the WSe$_2$ surface decreases and needs to be refreshed.
This effect occurs faster for illumination at the exciton resonance.

\subsection*{Finite-Difference Time Domain Simulations}
The simulation environment spans a volume of \SI{25}{\micro\meter}$\times$\SI{3}{\micro\meter}$\times$\SI{4}{\micro\meter} with an uniform \SI{20}{\nano\meter} $\times$ \SI{20}{\nano\meter} $\times$ \SI{100}{\nano\meter} mesh grid in the region of the waveguide.
The mesh step in z-direction is quite large to reduce simulation size and time.
We ensured its validity by reproducing the results of exemplary simulations with $\Delta$z = \SI{4}{\nano\meter}.
The environment is surrounded by perfect-matching layer boundaries and the total run time per simulation is \SI{200}{\femto\second}.
The WSe$_2$ waveguide sits on top of a \SI{21}{\nano\meter} thin Indium-Tin Oxide film and a glass substrate filling the simulation environment in the negative z-direction. 
Its excitation edge is located at x = \SI{-4}{\micro\meter} with propagation in the positive x-direction.
All materials fill the environment in the y-direction.
The source is given by a Gaussian beam with a waist radius of \SI{5}{\micro\meter} at the excitation edge of the waveguide.
Its spectrum is centered around \SI{800}{\nano\meter} with a full width at half-maximum of \SI{100}{\nano\meter} and a temporal width of about \SI{16}{\femto\second}.
A 2D-field monitor located at the top surface of the WSe$_2$ waveguide records the electric field components in the wavelength range of \SIrange{650}{950}{\nano\meter}.

\subsection*{Atomic Force Microscope}
We use a commercial atomic force microscope (Veeco, Nanoscope V) and tip (Budgetsensors, TAP 150).
The AFM is installed on an anti-vibration table.
The aluminum coated tip has a resonance frequency of about \SI{150}{\kilo\hertz} and we measure the height of the WSe$_2$ flakes in tapping mode.

\section*{Data Availability Statement}
The data that support the findings of this study are openly available in Zenodo at \url{http://doi.org/10.5281/zenodo.14168816}.

\section*{Author information}
\subsection*{Author contributions}

M.B. conceived the ideas and designed the experiments.
T.E. carried out the PEEM experiments, the FDTD simulations and is responsible for the data evaluation.
F.D. and K.R. synthesized the \wse{} crystals.
M.S. performed the mechanical exfoliation of the \wse{} flakes onto the ITO substrates.
V.D.G. carried out the AFM measurements.
T.E. and M.B. wrote the manuscript.

\subsection*{Funding}
The PEEM study and the sample growth was supported by the German Research Foundation (DFG) through Projects 499426961 and 434434223 (CRC 1461), respectively.

\section*{Acknowledgements}
The authors acknowledge Prof. Dr. Nahid Talebi for discussions regarding the thickness dependent coupling and the comparison of the PEEM and CL data.

\section{Supporting Information}
Supporting Information Available: 
Additional results measured in the 3PPE regime and additional simulations of the electric field components, thickness dependent dispersion and the field localizations in the waveguide.

\providecommand{\latin}[1]{#1}
\makeatletter
\providecommand{\doi}
  {\begingroup\let\do\@makeother\dospecials
  \catcode`\{=1 \catcode`\}=2 \doi@aux}
\providecommand{\doi@aux}[1]{\endgroup\texttt{#1}}
\makeatother
\providecommand*\mcitethebibliography{\thebibliography}
\csname @ifundefined\endcsname{endmcitethebibliography}  {\let\endmcitethebibliography\endthebibliography}{}

\end{document}


\title{Supplemental Material}
\subtitle{Photoemission electron microscopy of exciton-polaritons in thin \wse{} waveguides}
\author{T. Eul$^{1}$, M. Sabir$^{1}$, V. DeManuel-Gonzalez$^{1}$, F. Diekmann$^{1}$, K. Rossnagel$^{1,2}$ and M. Bauer$^{1,2}$}
\publishers{1 Institute of Experimental and Applied Physics, Kiel University, 24098 Kiel, Germany\\2 Kiel Nano, Surface and Interface Science KiNSIS, Kiel University, 24118 Kiel, Germany}
\date{}
\maketitle
\tableofcontents
\listoffigures

\section{Three-Photon Photoemission and beam damage}

\begin{figure}[H]
    \centering
    \includegraphics{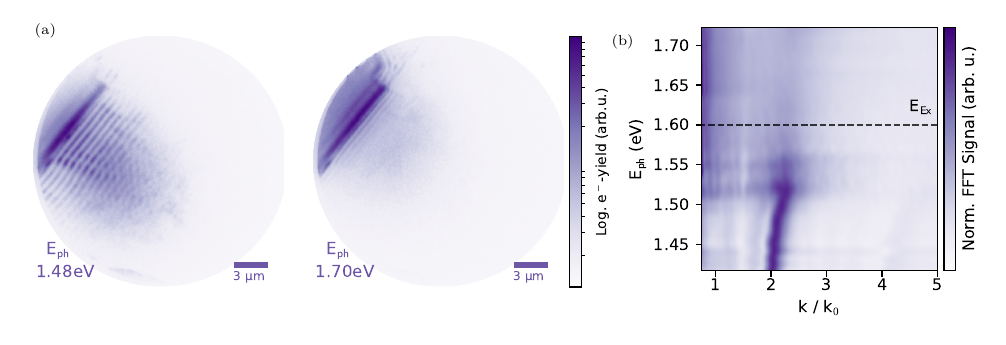}
    \caption[Dispersion with three-photon photoemission]
    {
    \textbf{Dispersion measurement in the three-photon photoemission regime.}
    (a) PEEM images recorded with photon energies of \SI{1.48}{\electronvolt} and \SI{1.70}{\electronvolt} with a polarization parallel to the excitation edge.
    The color scale is logarithmic to enhance the contrast of the comparibly low 3PPE yield.
    (b) Dispersion plot extracted from the PEEM images measured over the range of \SIrange{720}{875}{\nano\meter}.
    }
    \label{fig:3PPE}
\end{figure}

Since it is possible in our setup to focus the laser beam size down to \SI{5}{\micro\meter} on the sample surface, we initially forewent the lowering of the work function of the sample via the potassium evaporation.
The tight focus allowed us to image the wave patterns on the \wse-flakes in the three-photon photoemission (3PPE) regime, despite the high work function.
\fig{3PPE} shows an example of a dispersion measurement on a second WSe$_2$-flake without potassium on the sample.
For this series, we used laser powers in the range of \SIrange{50}{100}{\milli\watt} depending on the central wavelength from \SIrange{875}{720}{\nano\meter}.
The color scale in \fig{3PPE}(a) is logarithmic.
The dispersion of the lower polariton branch in \fig{3PPE}(b) is very similar to the one in Fig.~4 of the main text corresponding to a flake with a thickness of \SI{30}{\nano\meter}.
While we are able to resolve the upper polariton branch with the 2PPE experiment described in the main text, we cannot do so in the 3PPE regime due to the stronger influence on the absorption.
In general, the overall longer exposure times required to compensate for the drastically lower 3PPE-yield, the reduced area of photoemission yield and a lower contrast to the camera background noise make the experiments rather challenging.
Additionally, we were always operating at the threshold to beam damage of the WSe$_2$-flakes, especially when measuring with the photon energies close to the exciton resonance.

\begin{figure}[!b]
    \centering
    \includegraphics{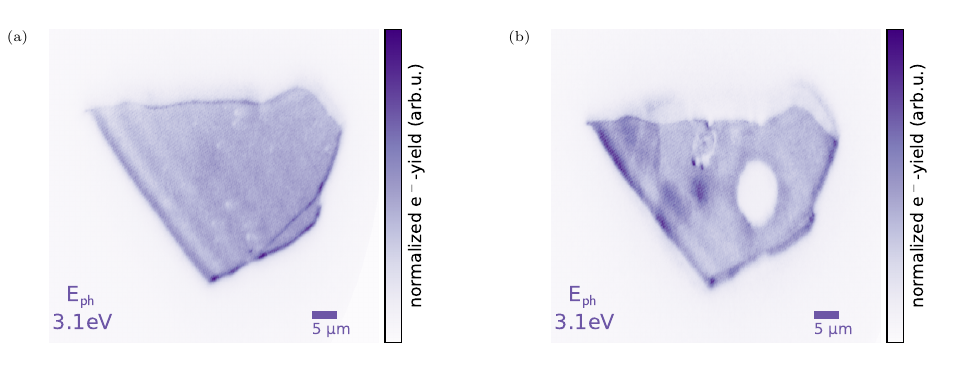}
    \caption[Beam damage with three-photon photoemission]
    {
    \textbf{Beam damage with three-photon photoemission.}
    PEEM images of a WSe$_2$ flake recorded with \SI{400}{\nano\meter}, the second harmonic of our laser, under grazing incidence illumination (\SI{65}{\degree}). 
    (a), (b) Images before and after normal incidence illumination of the flake with the fundamental of the laser at \SI{800}{\nano\meter}, which burned a hole through the flake.
    }
    \label{fig:Burning}
\end{figure}
\fig{Burning} showcases an example of beam damage of a WSe$_2$ flake recorded with \SI{400}{\nano\meter}, the second harmonic of our laser, under grazing incidence illumination (\SI{65}{\degree}).
The images in \fig{Burning} were taken before (a) and after (b) illumination with the fundamental of the laser at \SI{800}{\nano\meter}.
The bright spot indicates that in this area a hole was burned through to the substrate. In the shown example, this was achieved even with the laser not being mode-locked, i.e. in continuous wave mode.
Other areas of the flake show a slightly higher emission yield than before, which suggests the surface has been altered without burning through yet.
These results lead us to resort to the potassium evaporation to lower the work function, which allowed us to measure the samples in the two-photon photoemission (2PPE) regime.
Consequently, we could increase our laser spot size to increase the photoemission yield area over the entire field of view and we minimized the potential for beam damage.

\section{Electric field components of the TE- and TM-mode} \label{sec:components}

\begin{figure}[!b]
    \centering
    \includegraphics{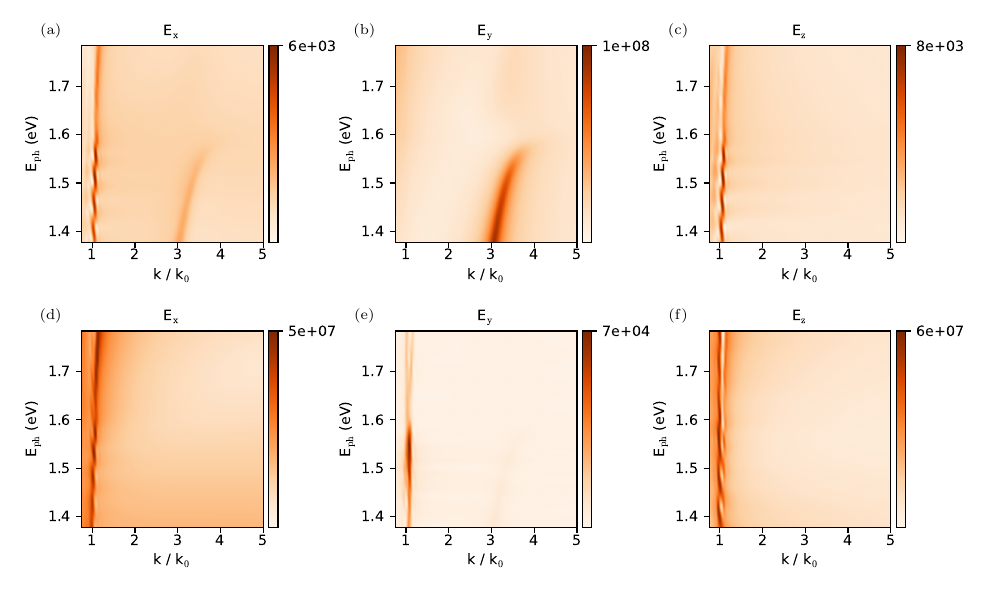}
    \caption[Electric field components for the TE- and TM-mode]
    {
    \textbf{Electric field components for the TE- and TM-mode.}
    Photon energy dependent Fourier amplitudes of the individual electric field components simulated for a thickness of \SI{55}{\nano\meter}.
    (a-c) $E_x, E_y, E_z$ of the TE-mode with highest amplitude in the $E_y$-component.
    (d-f) $E_x, E_y, E_z$ of the TM-mode with highest amplitude in the $E_x$- and $E_z$-component.
    }
    \label{fig:Components}
\end{figure}

This section showcases the individual electric field components extracted from the Finite-Difference Time-Domain (FDTD) simulations shown in Fig.~3 of the main text.
In this case, the WSe$_2$ film has a thickness of \SI{55}{\nano\meter}.
The top and bottom of \fig{Components} show the simulation results for the $E_x, E_y, E_z$ components for a polarization parallel and perpendicular to the excitation edge, respectively.
The excitation edge of the flake is oriented along the y-direction, the surface normal is in the z-direction and the waveguide mode propagates in the x-direction.
The number next to the color bar represents the relative maximum Fourier amplitudes of the individual electric field components.
In the case of an excitation of the modes with a polarization parallel to the edge (top row), the majority of the intensity resides in the $E_y$ component.
This component is parallel to the edge and perpendicular to the propagation direction of the mode.
Hence, the mode conforms to the transverse electric (TE) mode.

In the other case, with a polarization of the light perpendicular to the edge, the majority of the intensity is in the $E_x$ and $E_z$ components.
This is expected for the transverse magnetic (TM) mode, where only the magnetic field is perpendicular to the propagation direction.
Since the waveguide mode does not propagate in a straight line in the x-direction, as it gets reflected at the top and bottom interfaces in a zig-zag motion, its electric field gets projected onto the x- and z-direction.
The field component $E_x$ in (d) shows a feature with slightly higher k values than the light line for higher photon energies, which we attribute to this TM-mode.
At lower photon energies it is indistinguishable from the light line.

The $E_z$ component on the other hand shows two features.
Firstly, a line following the vacuum light dispersion, which emerges when the simulation source hits the edge since the edge allows coupling in any direction.
This electric field is mostly located outside of the waveguide after coupling (see \sct{confinement}), which is why it does not get influenced by the refractive index of the WSe$_2$.
Secondly, we can see the same feature that is visible in the $E_x$-component, albeit with slightly lower amplitude.
The difference in amplitude of this feature between the $E_x$- and $E_z$ component stems from the different projection of the field on the two directions for this particular thickness.
For lower energies again the two features are not distinguishable and a mixing of the two features occurs.

\section{Thickness dependence of the exciton-polariton coupling} \label{sec:confinement}

For both TE- and TM-modes, the coupling to the exciton in the WSe$_2$ waveguides occurs after reaching a threshold thickness of the waveguide.
The thickness dependent simulations of the mode confinement and the dispersion in this section highlight this threshold.
The general argument is that enough of the electric field has to be localized within the waveguide for coupling to the exciton to occur.

\subsection*{TE-mode}

\begin{figure}[!b]
    \centering
    \includegraphics{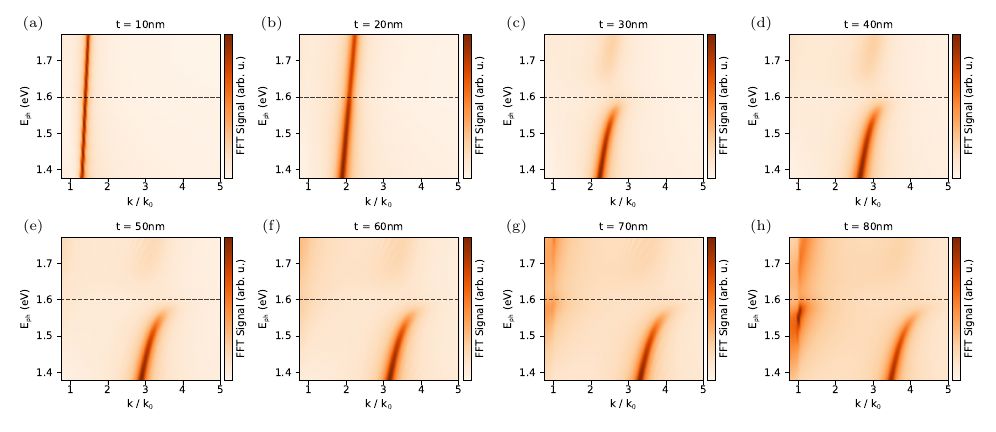}
    \caption[Thickness dependence of the dispersion for the TE-mode]
    {
    \textbf{Thickness dependence of the dispersion for the TE-mode.}
    Simulated dispersion relations for the TE-mode for waveguide thickness $t$ in the range of \SIrange{10}{80}{\nano\meter}.
    }
    \label{fig:Thickness_TE}
\end{figure}

Starting with the TE-mode, \fig{Thickness_TE} shows the simulated dispersions for waveguide thicknesses $t$ in the range of \SIrange{10}{80}{\nano\meter}.
Overall the absolute k-values shift towards higher momenta with increasing thickness as expected for a planar waveguide.
For $t= $\SI{80}{\nano\meter} the second TE mode starts to emerge at low momenta.
More interestingly, we observe no coupling to the exciton for $t= $\SI{10}{\nano\meter} and $t= $\SI{20}{\nano\meter} but coupling occurs for $t= $\SI{30}{\nano\meter}.
The coupling threshold obviously lies between $t= $\SI{20}{\nano\meter} and $t= $\SI{30}{\nano\meter} for the TE-mode.

\begin{figure}[t]
    \centering
    \includegraphics{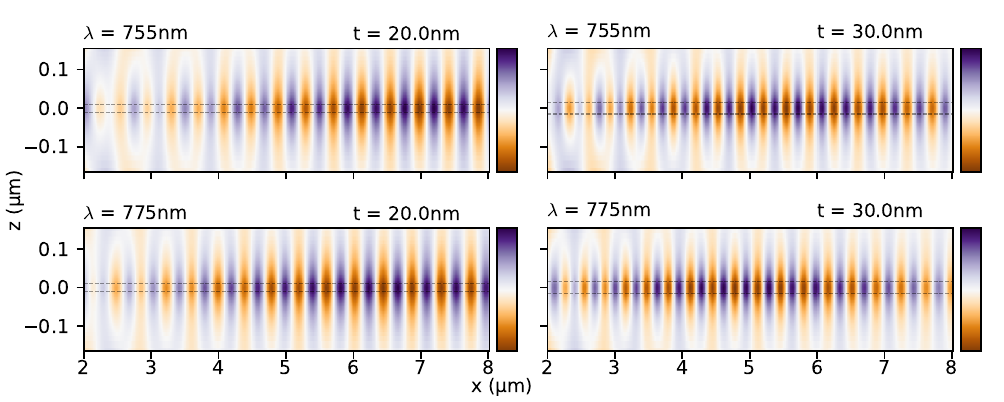}
    \caption[TE-mode confinement in the waveguide - $E_y$]
    {
    \textbf{TE-mode confinement in the waveguide}.
    $E_y$-component of the TE-mode simulated for \SI{755}{\nano\meter} (top) and \SI{775}{\nano\meter} (bottom) as well as for $t= $\SI{20}{\nano\meter} (left) and $t= $\SI{30}{\nano\meter} (right).
    The fields have been Fourier filtered in the direction of the waveguide to remove the strong intensity drop off caused by the focus of the source.
    }
    \label{fig:Confinement_TE}
\end{figure}

To explain this threshold, we extracted the mode localization from the simulations for different thicknesses.
For this, rather than using an in-plane monitor in the xy-plane as before, we set the monitor in the middle of the waveguide in the xz-plane.
\fig{Confinement_TE} shows the extracted electric fields $E_y$ of the TE-mode for the two wavelengths \SI{755}{\nano\meter} and \SI{775}{\nano\meter} as well as the two thicknesses $t= $\SI{20}{\nano\meter} and $t= $\SI{30}{\nano\meter}.
We chose these wavelengths since they are close to the exciton resonance. 
To remove the strong intensity drop off caused by the focus of the source, we extracted the fields away from the edge, starting at \SI{2}{\micro\meter} and applied a high pass Fourier filter along the direction of the waveguide.
The dashed horizontal lines indicate the waveguide boundaries in the z-direction.

All images in \fig{Confinement_TE} show that the electric field maxima of the TE-mode are all localized within the waveguide but that the electric field also extends considerably beyond the waveguide boundaries.
Remarkably, the localization of the field is less pronounced for the smaller thickness.
Since the waveguide is simply too thin at some point (presumably between \SIrange{20}{30}{\nano\meter}), the extension outside of the waveguide means that the field amplitude within the interaction volume with the exciton becomes too small to support coupling.

\subsection*{TM-mode}

\begin{figure}[H]
    \centering
    \includegraphics{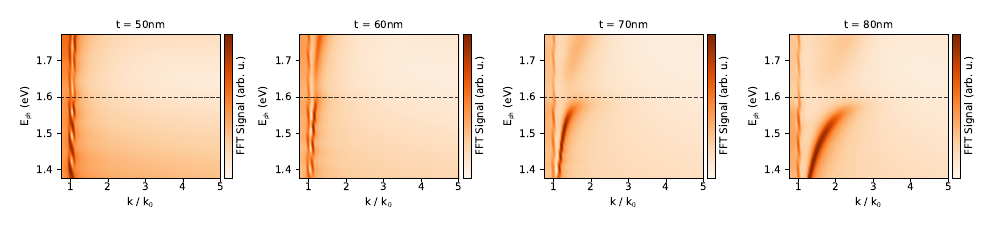}
    \caption[Thickness dependence of the dispersion for the TM-mode]
    {
    \textbf{Thickness dependence of the dispersion for the TM-mode}.
    Simulated dispersion relations for the TM-mode for thickness $t$ in the range of \SIrange{50}{80}{\nano\meter}.
    }
    \label{fig:Thickness_TM}
\end{figure}

\begin{figure}[t]
    \centering
    \includegraphics{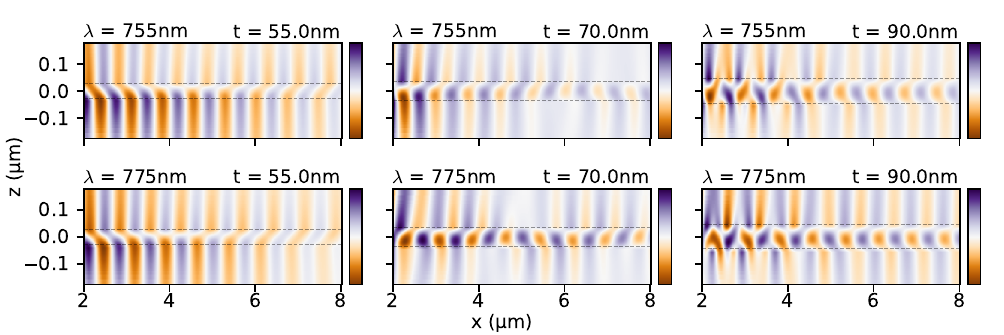}
    \caption[TM-mode confinement in the waveguide - $E_x$]
    {
    \textbf{TM-mode confinement in the waveguide}.
    $E_x$-component of the TM-mode simulated for \SI{755}{\nano\meter} (top) and \SI{775}{\nano\meter} (bottom) as well as for $t= $\SI{55}{\nano\meter}, $t= $\SI{70}{\nano\meter} and $t= $\SI{90}{\nano\meter} (from left to right).
    The fields have been Fourier filtered in the direction of the waveguide to remove the strong intensity drop off caused by the focus of the source.
    }
    \label{fig:Confinement_TMx}
\end{figure}

\fig{Thickness_TM} shows the simulated dispersions for the TM-mode for $t$ in the range of \SIrange{50}{80}{\nano\meter}.
Coupling to the exciton starts to occur from $t=\SI{60}{\nano\meter}$, at least with a slight bend in the dispersion.
At \SI{70}{\nano\meter} the coupling is very evident.
This threshold between \SI{50}{\nano\meter} and \SI{60}{\nano\meter} can again be explained with the localization of the electric field within the waveguide:
\fig{Confinement_TMx} shows the electric fields of the $E_x$-component of the TM-mode again for the two wavelengths of \SI{755}{\nano\meter} (top) and \SI{775}{\nano\meter} (bottom) close to the exciton resonance.
For this case, we take a look at the three thicknesses, $t= $\SI{55}{\nano\meter}, $t= $\SI{70}{\nano\meter} and $t= $\SI{90}{\nano\meter}.
In the two latter cases, we can see a clear confinement of the mode within the waveguide for both wavelengths. This is the reason why exciton-polariton coupling can occur.
For the lower thickness however, the field amplitude is much less confined and it has equal amplitudes in the waveguide and in the region below the waveguide.
Just like in the case of the TE-mode, the electric field is not strong enough within the interaction volume to support exciton-polariton coupling for lower thicknesses.

\begin{figure}[H]
    \centering
    \includegraphics{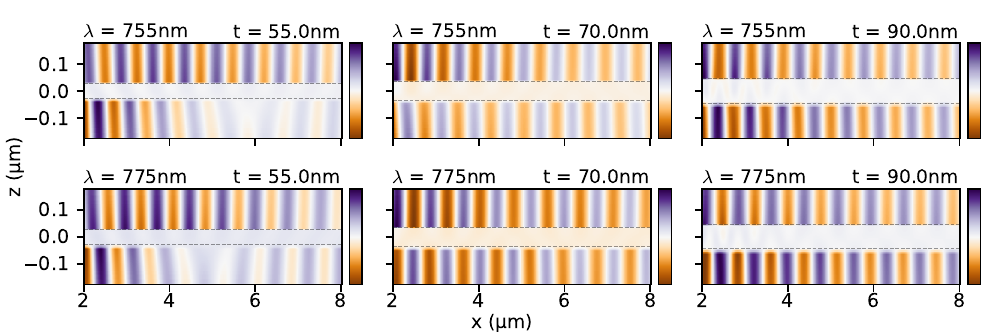}
    \caption[TM-mode confinement in the waveguide - $E_z$]
    {
    \textbf{TM-mode confinement in the waveguide}.
    $E_z$-component of the TM-mode simulated for \SI{755}{\nano\meter} (top) and \SI{775}{\nano\meter} (bottom) as well as $t= $\SI{55}{\nano\meter}, $t= $\SI{70}{\nano\meter} and $t= $\SI{90}{\nano\meter} (from left to right).
    The fields have been Fourier filtered in the direction of the waveguide to remove the strong intensity drop off caused by the focus of the source.
    }
    \label{fig:Confinement_TMz}
\end{figure}

For completion, \fig{Confinement_TMz} shows the field localization for the $E_z$-component of the TM-mode.
As already alluded to in \sct{components}, in this case the field is mostly localized outside of the waveguide.
